%% file: three.tex
\documentstyle[preprint,floats,aps,epsf,prb]{revtex}

\input defs

\input defs2

\def\joinrelde{\mathrel{\mkern-11mu}}
\def\joinrel{\mathrel{\mkern-9mu}}

\def\joinrelw{\mathrel{\mkern-6mu}}

\def\relbd{\mathrel{{\bf\smash{{\phantom- \above1pt \phantom-
}}}}}
\def\ltdash{\raise-1.8pt\hbox{$\scriptscriptstyle |$}}

\def\vk{{\rm  \vec k}}

\def\sl{\it}

\catcode`\@=11       
\def\listzigurename{List of Zigures}

\def\zigurename{Zigure}

\def\listofzigures{\section*{\listzigurename
    \@mkboth{\uppercase{\listzigurename}}{\uppercase{\listzigurename}}}%
  \@starttoc{lof}}

\def\l@zigure{\@dottedtocline{1}{1.5em}{2.3em}}

\let\l@table\l@zigure

\newcounter{zigure}
\def\thezigure{\@arabic\c@zigure}

\def\fps@zigure{tbp}
\def\ftype@zigure{1}
\def\ext@zigure{lof}
\def\fnum@zigure{\zigurename~\thezigure}
\def\zigure{\@float{zigure}}
\let\endzigure\end@float
\@namedef{zigure*}{\@dblfloat{zigure}}
\@namedef{endzigure*}{\end@dblfloat}
\catcode`\@=12
\begin{document}
\draft

\preprint{6-3-94}
\title{
Three Body bound-states and the development of odd frequency 
pairing}
\author{P. Coleman$^1$, E. Miranda $^1$ and A. Tsvelik $^{2}$}
\address{
$^1$ Serin Physics Laboratory,
Rutgers University,
PO Box 849,
Piscataway NJ, 08855.
\\
$^2$ Dept. of Physics,
Oxford University, 1 Keble Road, Oxford OX1 3NP, UK.
}
\maketitle
\begin{abstract}
We propose that the development of odd-frequency superconductivity is
driven by the formation of neutral three body bound-states.  Using a
three-body bound-state ansatz we develop a mean-field theory for
odd-frequency pairing within the Kondo Lattice model.  Three body
bound-state formation leads to the formation of a gapless band of
fermions with a neutral, spinless Fermi surface.  We
discuss the low energy excitations of these modes, suggesting them as
a possible explanation for the absence of anisotropy in the thermal
conductivity of heavy fermion superconductors.
\end{abstract}

\vskip 0.2 truein

%
%
\onecolumn
Although three body fermion states are ubiquitous in many branches of
particle, nuclear and atomic physics, there has been very little study
of their impact on collective condensed matter behavior.
In this paper, we propose that three body bound-state 
instabilities\cite{ruckenstein} can
precipitate a phase transition, and we identify the broken symmetry state
that forms as an odd-frequency superconductor.  Berezinskii
\cite{berezinskii} originally proposed that superconductivity could
occur through the development of a state with an odd frequency gap
function
\begin{equation}
\ul{\Delta}(\om, k) = -\ul{\Delta}(-\om, k)
\end{equation}
Unlike conventional superconductivity, there is no weak-coupling
Cooper instability associated with this type of state.\cite{abrahams}
Indeed the proper identification of an appropriate class of
many body instability for odd-frequency pairing
has proved a major obstacle to concrete theoretical
realization of Berezinskii's proposal.

Our particular interest in odd-frequency pairing is motivated by heavy
fermion superconductors,\cite{hf,sauls} where superconductivity appears
to involve the active participation of local moments in the condensation
process.  Vanishing coherence factors particular to odd frequency
pairing may offer a means to reconcile the large linear specific heat
and isotropic thermal conductivity observed in these superconductors
with the absence of a linear Korringa relaxation in the NMR
data.\cite{previous,cv,louis,upt3nmr,ube13nmr} In this paper we show how
the development of such hypothetical states can be directly associated
with the formation of three-body bound-states between electrons and
local moments.

A three-fermion bound-state is a composite fermion where selection
rules  stabilize it against decay into its 
constituents.
A good example  is the  He-3 atom: 
a neutral bound-state formed from a $S=1/2$ nucleus and
two electrons. Since the low-energy dynamics of the atom
involve
rigid center-of-mass motion,  correlation functions
of the bound  fermions are determined by their factorization
into triplet contractions
\begin{equation}
\mathrel{\mathop
{{
 \hat\psi_{\up}({\bf 1})
\hat \psi_{\dw}({\bf 2})
 {\cal N}_{\si}({\bf 3})
}}
^{\dsp \ltdash
\joinrel\relbd
\joinrelw\relbd
\joinrelw\relbd
\joinrelw\relbd\joinrel
 \ltdash
\joinrelde\relbd
\joinrel\relbd
\joinrel\relbd
\joinrel\relbd
\joinrel\relbd\joinrel
 \ltdash
{\phantom{_{\si}(3)}}
}}
=\int
\Lambda({\bf 1},{\bf 2},{\bf 3};x)
\hat \Phi_{\si}(x) dx 
\end{equation}
where $\Phi\dg (x)$ creates the He-3 fermion, 
$\psi$ represents the electron fields, ${\cal N}$ the nucleus,
and $\Lambda$ is the atomic wave function of the He-3 atom. 
It is this factorization
into  ``three body contractions'' which enables us
to replace  the composite operators by a single fermion $\Phi(x)$
in the low energy physics of He-3.

Can such bound-states
form {\sl collectively} in an electronic system ?
A well-known theorem of Yang \cite{yang} precludes
the development of any literal off-diagonal long-range order
of a three-body fermion. Here we propose an alternative
possibility, where two electrons and a hole bind to form
a   neutral fermion through the collective formation
of a three-body condensate:\cite{montorsi}
\begin{equation}
\mathrel{\mathop{{
\psi_{\alpha}\dg(1) \psi_{\beta}(2) \psi_{\gamma}(3)}}
^{\dsp \ltdash
\joinrel\relbd
\joinrel\relbd
\joinrel\relbd
\joinrel\relbd
\joinrelw\relbd\joinrel
 \ltdash
\joinrelde\relbd
\joinrel\relbd
\joinrel\relbd
\joinrel\relbd\joinrel
 \ltdash
{\phantom{_{\si}(3)}}
}}
= \int 
\Lambda_{\alpha \beta \gamma}(1,2,3; x) \hat \phi (x)dx
\label{the_bare_essentials}
\end{equation}
Here 
$\phi(x)= \phi\dg(x)$ represents the neutral bound-state
fermion.
$\Lambda_{\alpha\beta \gamma}$  is a  {\sl complex} wave function 
which carries
the phase information associated with the charge of the
bound electrons, playing the role of a collective variable.
We shall argue that if the phase of the three-body wave function develops a
rigidity, an odd frequency superconductor is formed.
For simplicity we  consider a wave function  that 
is symmetric in electron co-ordinates  $2$ and $3$, and hence
antisymmetric in the corresponding spin variables
\begin{equation}
\Lambda_{\alpha \be \gamma} (1,2,3;x) = \si^2_{ \beta \gamma} \si^2_{ \alpha 
\eta} A_{\eta}(1,2,3;x)
\end{equation}		
where $\si^2$ is a Pauli matrix.
The coarse-grained average 
$A(x)\equiv \int_{1,2,3}{A(1,2,3;x)}$ plays the role of
a  {spinorial order parameter}.\cite{note}
By contracting the  spin-indices, we 
see that $A$ describes
the binding between spins and  electrons,
\begin{equation}
\mathrel{\mathop{{
[ \vec S(1)\cdot
\vec \si_{\alpha \beta} 
] 
\psi_{\be}(2) }}
^{\dsp \ltdash
\joinrel\relbd\joinrel
 \ltdash
\joinrelde\relbd
\joinrel\relbd
\joinrel\relbd
\joinrel\relbd\joinrel\relbd
\joinrel\relbd
\joinrel\relbd
\joinrel\relbd\joinrel
 \ltdash
{\phantom{_{\be}(2)}}
}}= \int dx A_{\alpha}
(1,1,2;x)\hat \phi(x)
\end{equation}
where $\vec S= {1 \over 2}\psi\dg \vec \si \psi $ is the spin density.

A specific model for this type of bound-state instability is provided
the ``Kondo lattice'' Hamiltonian  
\begin{eqnarray}
H= \sum_{\vk} \eps_{\vk} \psi\dg_{\vk}\psi_{\vk}
+ \sum_j H_{int}[j]\label{strip}
\end{eqnarray}
where
$\psi\dg_{\vk}$ is a conduction electron spinor, coupled to an array
of $S=1/2$ local moments $\vec S_j$ 
via an antiferromagnetic exchange interaction
\begin{eqnarray}
H_{int}[j]=
J(\psi\dg_{j} \vec \si \psi_{j}) \cdot \vec S_j.
\label{kondo}
\end{eqnarray}
where $\psi_j$ denotes the conduction electron in a tight binding representation. This Hamiltonian is of particular interest
as a toy model for heavy fermion metals.
An electron scattering off a magnetic ion 
at site $j$ couples directly
to the three-body spinor
\begin{equation}
\xi_{j\alpha}  = (
\vec S_j \cdot \vec \si_{\alpha \beta})
 \psi_{j \beta}
\label{spino}
\end{equation}
Irreducible Green functions of $\xi$ determine
the self-energies of the conduction electrons.
In a superconducting
state,  the normal and anomalous 
components of the electron self-energies are given by 
\begin{eqnarray}
\begin{array}{rl}
\Sigma(\ka) =& \dsp J^2\la\la \xi ( \ka) \xi \dg(\ka) \ra\ra \cr
\Sigma_A(\ka) =& \dsp J^2\la\la \xi ( \ka) \xi (-\ka)\ra\ra \cr
\end{array}
\label{selfen}
\end{eqnarray}
where double brackets denote one-particle irreducibility with respect to
conduction electrons and $\ka \equiv( \om, \vk)$. 
The development of bound-state resonances
associated with $\xi$ can thus severely modify the electronic
properties of the lattice.

To provide a controlled treatment of 
fermionic bound-states
in the Kondo lattice model, we  develop a $1/N$ expansion within
a class
of ``$O(N)$ Kondo models''.
In this generalization, the local anticommuting algebra
of the original spin-1/2 operators is preserved
\begin{equation}
\begin{array}{rl}
\{\si^{\mu}, \si^{\nu}\} =& \dsp 2 \delta_{\mu \nu}\cr
\{S_j^{\mu}, S_j^{\nu}\} =& \dsp {1 \over 2} \delta_{\mu \nu}\cr
\end{array}
\end{equation}
but the number of components of each operator is generalized from
three in the model of interest, to an arbitrary odd number $N$.  
This is the algebra of $N$ component gamma matrices. For general $N$, 
the conduction
electron spinors have dimension $2^{[N/2]}$, where
$[\ ]$ denotes the integer part. 
For all $N$, the Kondo interaction Hamiltonian can be
written
\begin{eqnarray}
H_{int}[j]=
-{ \tilde J \over  N}(\xi\dg_{j} \xi_{j })
\label{kondo2}
\end{eqnarray}
where we replace $J= \tilde J /N$. 

We now calculate the correlation function of $\xi$ in the 
normal state, in a scheme that is exact to
order $O(1/N)$.  Since $\xi$ is a fermion,
a fluctuating component of the operator
must survive development of bound-state
singularities  in its correlation functions. 
To describe these intrinsic fluctuations
we introduce the device of
a ``spectator fermion''  ($\phi_j$): a 
neutral fermion $\phi_j=\phi_j\dg$, defined at each site which 
which
satisfies canonical anticommutation rules, 
$\{\phi_j,\phi_k\}=\delta_{jk}$, commuting with the spin operators
and anticommuting with all
other fermions.
$\phi_j$, so defined, commutes with the Hamiltonian and
decouples from the dynamics; its 
correlation functions display intrinsic
fluctuations 
\begin{equation}
\la {\rm T}\phi_{j}(1) \phi_{k}(2)\ra
= {\delta_{jk} \over 2}{\rm sgn}(\tau_1-\tau_2)
\end{equation}
We construct a  ``bosonic partner'' to the three-body operator
$\xi$ as follows
\begin{equation}\hat A_j
= 2 \phi_j\xi_j
.
\label{definer}
\end{equation}
Since
$\phi_j^2={1\over 2}$, we may  invert this relation to obtain $
\xi_j =\phi_j \hat A_j $. 
The spectator fermion
factors out of the dynamics, so we may relate the correlation functions
of $\hat A_j$ and $\hat \xi_j$, 
\begin{eqnarray}
\begin{array}{rl}
 \la \hat A_j(\tau) \hat A\dg_k(0)\ra = \dsp{ 2}{\rm sgn}(\tau) \delta_{jk}\la\la \xi_j(\tau) \xi_j\dg (0)\ra\ra
\end{array}
\label{fermicond2}
\end{eqnarray}
We can also reversibly transform the spin operators
into {\sl fermionic} counterparts
\begin{equation}
\vec \eta_j=2\phi_j\vec S_j
\label{kondo7}
\nonumber
\end{equation}
which satisfy 
$
\{\eta^a[j],\eta^b[k]\}
= \delta^{ab}\delta_{jk}
$ and are {\sl real} $\eta^a[j]=\eta^{a \dagger}[j]$.
The canonical commutation properties of these ``Majorana''  spin fermions 
permits us to use diagrammatic methods
without invoking a Gutzwiller projection.
\cite{details}
Since
$
\hat A_j = [ \vec \eta_j \cdot \vec \si ] \psi_j$, 
the original interaction can be written
\begin{equation}
H_{int}[j] = - { \tilde J \over 2 N} \hat A\dg_j \hat A_j\label{newform}
\end{equation}

Consider the diagrammatic expansion for the local
susceptibility
\begin{equation}
\la A_i(\tau) A\dg_j(0)\ra=\delta_{ij}\chi(\tau) 
\end{equation}
in powers of $\tilde J/N$.  (Fig. (1.))
The non-interacting
value  is 
$
\chi^0(\tau)
= {N\over 2 } { \rm sgn} ( \tau) {\cal G} ( \tau)
$
where 
$
{\cal G}(\tau) = 
\la \psi_i(\tau) \psi\dg_i(0)\ra 
$ is the local electron propagator. In the frequency domain
\begin{equation}
\chi^0(i\nu_n) = {N \over 2}\sum_{\vk}
{{\rm tanh}  (\beta \eps_{\vk}/2) \over \eps_{\vk}-i \nu_n }
\qquad\ (\nu_n = 2 \pi T n)
\end{equation}
is logarithmically divergent at low energies
$\chi^0(0)\sim N\rho ln[D/T]$, where $\rho$ is the conduction
electron density of states and D the bandwidth.
Each fermion loop contributes a factor $N$ and each interaction line
contributes a factor $1/N$, thus the leading order $O(N)$ contribution
to $\chi$ is given by the RPA sum
\begin{equation}\chi_{RPA}(i\nu_n) = N{{ \tilde \chi^0(i \nu_n)\over
( 1 - {\tilde J\over 2} \tilde \chi^0(i \nu_n))}},
\end{equation}

\begin{zigure}[here]
\epsfxsize=4.0truein
\hskip 1.0truein\epsffile{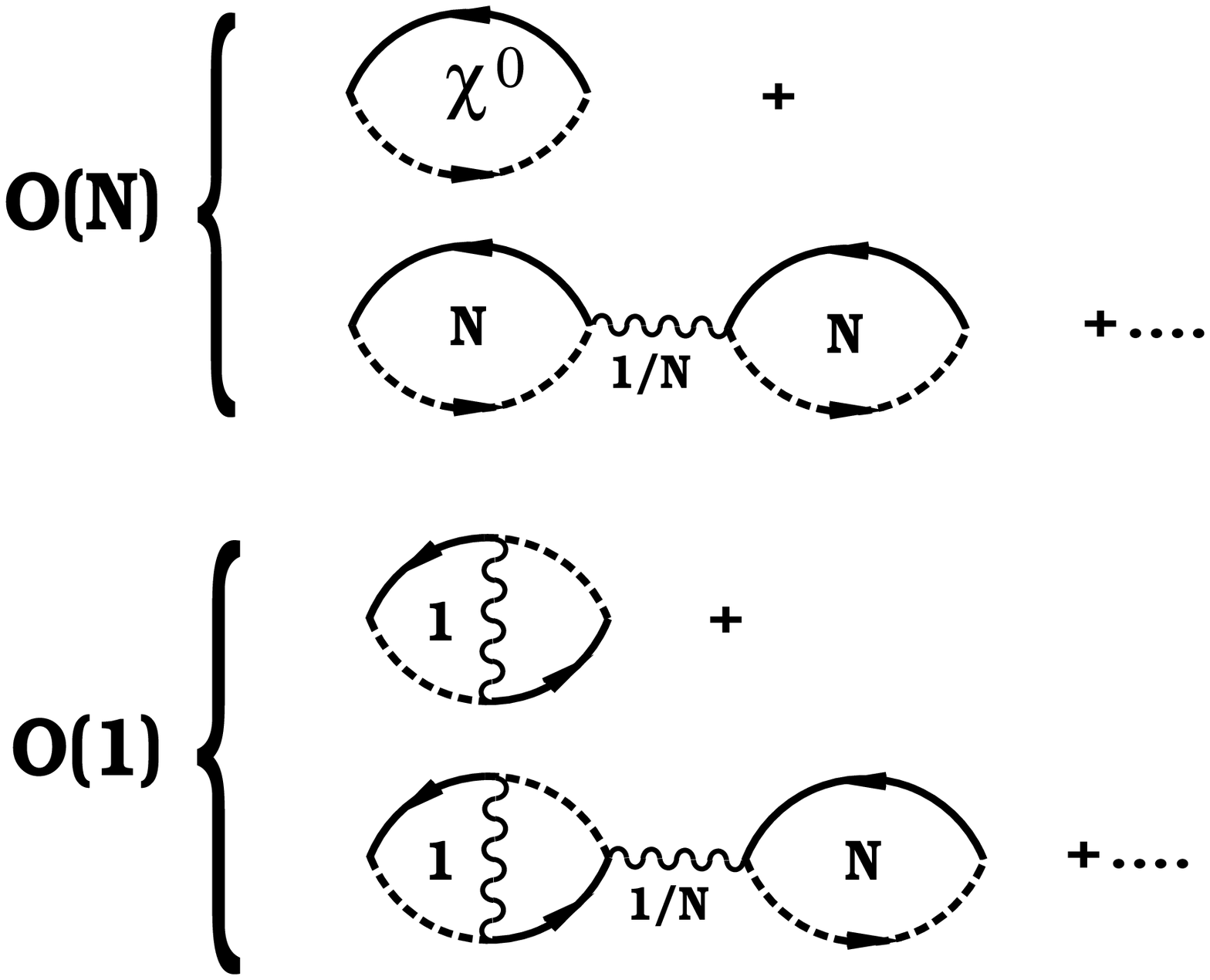}
\end{zigure}

\noindent{\bf Fig. 1. } 
Diagrammatic expansion of $\chi(\inu)$ in a $1/N$ expansion for the
$O(N)$ Kondo model: solid lines refer to conduction electron
propagators, dashed lines to spin fermions, and a wavy line denotes the
interaction vertex.

\vskip 0.3truein
\newpage
\noindent where $N\tilde \chi^0= \chi^0$. If we now transform back to
the corresponding
fermionic response function of 
$\xi$, we find
\begin{equation}
\la \xi_i (\tau) \xi_i\dg(0)\ra
= {1 \over 2} {\rm sgn}(\tau) 
\chi_{RPA}(\tau)
\end{equation}
Since $\chi^0(0)$ is logarithmically divergent, this 
function develops a singularity at a {\sl  finite}
temperature $T_c\sim D e^{-2/\rho \tilde J}$, where 
$1={\tilde J \over 2} \tilde \chi^0(0)$.
This  singularity signals the formation of
a three body bound-state.

We now apply our bound-state ansatz to the three body
spinor $\xi_j$, writing
\begin{equation}
\xi_{j}(t) =  \la A_{j}\ra  \hat \phi_j(t) + \delta \xi_j(t)
\label{ansatz}
\end{equation}
where the first term is the three body contraction of the 
operator.
This  ansatz implies long-time correlations of the three body operator
\def\joinrel{\mathrel{\mkern-9mu}}
\def\relbd{{{\bf\smash{{\phantom- \above 1pt \phantom-
}}}}}
\begin{eqnarray}
\begin{array}{rl}
\la\la \hat \xi_j(\tau) \hat \xi\dg_k(0)\ra\ra  &\dsp
{\mathop{\relbar\joinrel\relbar\joinrel\longrightarrow}_{\vert \tau \vert \rarrow \infty}}
{\delta_{jk}
 \over 2}
{\rm sgn} (\tau)A_j \otimes A_j\dg 
\\
\la\la \hat \xi_j(\tau) \hat \xi_k(0)\ra\ra &\dsp
{\mathop{\relbar\joinrel\relbar\joinrel\longrightarrow}_{\vert \tau \vert \rarrow \infty}}
{\delta_{jk} \over 2}{\rm sgn} (\tau)A_j \otimes A_j^{\rm T}
\\
\end{array}
\label{fermicondensate3}
\end{eqnarray}
For $N=3$, The Fourier transform of (\ref{fermicondensate3}), inserted into (\ref{selfen})
then gives
\begin{equation}
\begin{array}{rl}
\dsp\Sigma^A(\ka) = 
&\dsp
[V \otimes V^{\rm T} ]\la \phi(\ka) \phi(-\ka) \ra=
{4[V \otimes V^{\rm T} 
] \over \om}. \\
\end{array}
\label{ofreq}
\end{equation}
where
$V_j = {\tilde J\over 2N} \la A_j \ra$.
This  establishes the direct link between three body bound-state formation
and an odd frequency pairing amplitude that scales as $1/\omega$.

To develop the mean-field theory for the condensed phase, we substitute
the bound-state ansatz into the Kondo exchange interaction, so that
$H_{int}[j]\rarrow H_{int}[V_j]$, where
\begin{eqnarray}
H_{int}[V_j]=
{2}\biggl[
\bigl(\xi\dg_{j}\hat \phi_j V_j + 
({\rm H.C.})\bigr)
+ {V\dg_j V_j\over J}\biggr]
+ O(\delta \xi\dg \delta\xi)\nonumber
\end{eqnarray}
Using (\ref{kondo7}), 
$2\xi\dg \phi \equiv \psi\dg(\vec \si \cdot \vec \eta)$, thus 
\begin{eqnarray}
H_{int}[V_j]=
\biggl[\psi\dg_{j}(\vec \si\cdot \vec \eta_j)V_j + 
({\rm H.C.})
\biggr]
+ {V\dg_j V_j\over J}
\label{kondo33}
\end{eqnarray}
Remarkably, the fusion of the neutral bound-states with the 
the spin variables  transforms them into propagating spin fermions.
This form of 
mean-field theory was previously derived using
an abstract Majorana spin-representation of spin-$1/2$ operators.\cite{previous}

\def\psk{\psi_{j}}
\def\pskd{\psi\dg_{j}}
\def\pskdm{\psi\dg_{-j}}
\def\psd #1{\psi^{#1\dagger}_{j}}
\def\psu #1{\psi^{#1}_{j}}
Formation of a gapless band of neutral fermions follows
naturally from our original ansatz. 
If we write $V_j = {V\over \sqrt{2}} z_j$, where $z_j$ is a unit
spinor, and decompose
$\psk$ into four Majorana components
$
\psk = { 1 \over \sqrt{2}}\bigl[
\psu 0 + i \vec \psk \cdot \vec \si
\bigr]z_j$.
then the scalar component $\psi_o$ decouples from the 
the hybridization
term $H_{int}[V_j]=-i V \vec \psi_j \cdot \vec \eta_j$.
The vector 
components of the conduction sea hybridize with the spin fermions
to form a gap $\Delta_g \sim V^2 / D$.
Residual ``scalar'' components of the conduction sea thus split
off below the quasiparticle continuum to produce a  gapless band.(Fig. 2.)
Spin and charge operators are off-diagonal bilinears in this Majorana basis,
which ensures the neutrality of the gapless Fermi surface and leads
to coherence factors that vanish linearly with energy.\cite{previous}

\begin{zigure}[here]
\epsfxsize=4.0truein
\hskip 1.0truein\epsffile{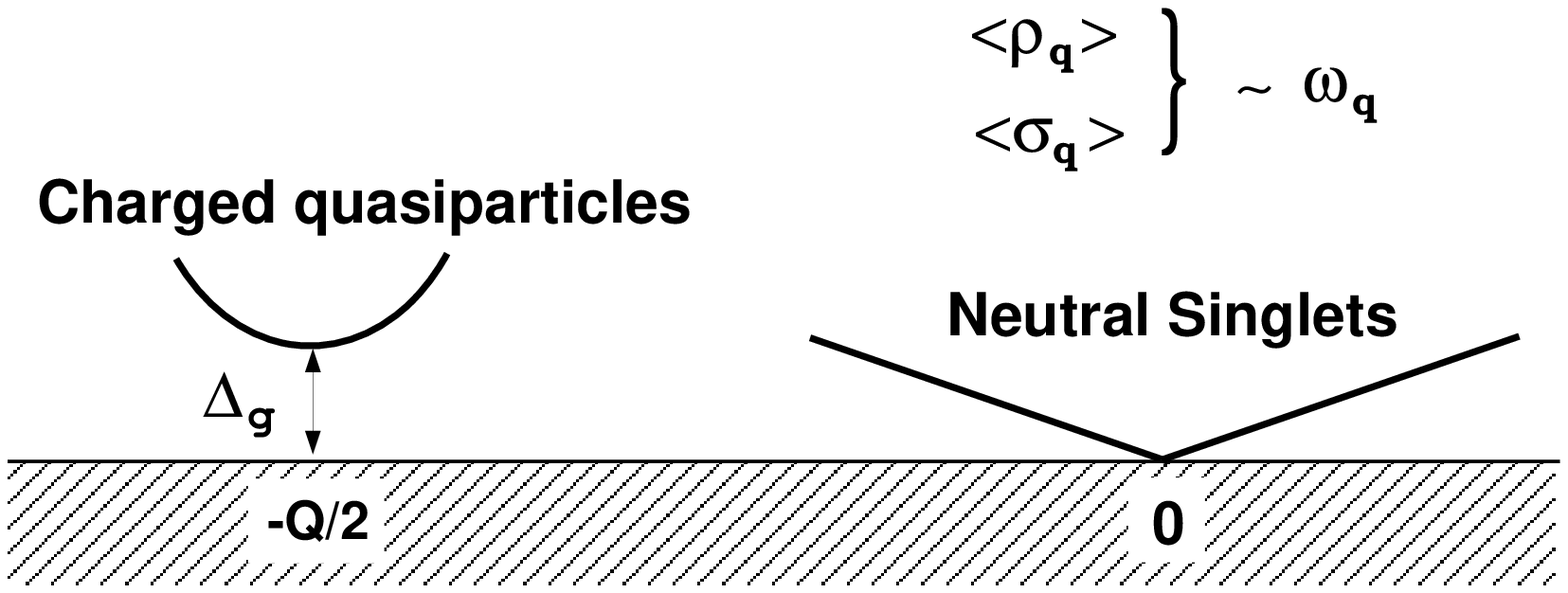}
\end{zigure}

\noindent{\bf Fig. 2.} Schematic illustration of the excitation spectrum predicted by the
mean-field theory.  Formation of the three body bound-state splits
a gapless band of neutral fermions off from the charged quasiparticle
continuum.

\vskip 0.3truein

The use of spectator fields 
differs 
fundamentally  from the more familiar  use of ``slave''
fields, since  the combined  Hilbert space of particles and spectators
{\sl replicates} the physical Hilbert space. 
There is no unwanted subspace 
with unphysical matrix elements to be projected out.
In a path
integral language the  decoupling of the spectator fields is associated
with a unique fermionic gauge invariance.
The appropriate Lagrangian is 
\begin{equation}
{\cal L} = \sum_{j}\{\psi_j\dg \partial_{\tau} \psi_j +{1 \over 2} \vec \eta_j \partial
\vec \eta_j + H_{int}[V_j]\}+H_c 
\end{equation}
where $V_j$ is  now a
fluctuating bosonic spinor field. This
Lagrangian is derived from the observation that the combined Hilbert space
of spectator fermions and spins is equivalent to the Hilbert
space of the Majorana spins: ${\rm Tr}_{\vec\eta}
\equiv {\rm Tr}_{\vec S, \phi }$.
The saddle point approximation to the path integral of this
Lagrangian corresponds to
the mean-field theory described  above. 
Suppose we couple a Grassmann source term $\alpha_j$ to the 
$\phi$ field
\begin{equation}
{\cal L} \rarrow {\cal L} + \sum_j \dot \alpha_j \phi_j,
\qquad\quad (\dot \alpha_j\equiv \partial_{\tau} \alpha_j).
\end{equation}
For $N=3$,  $\phi_j\equiv -2i \eta^x_j \eta^y_j \eta^z_j$.\cite{details}
We can always gauge away such a term by a transformation
\begin{eqnarray}
\begin{array}{rl}
\phi_j &\dsp
\rarrow \phi_j +  \alpha_j\\
\eta_j &\dsp
\rarrow \vec \eta_j + 2 \alpha_j \vec S_j\\
V_j &\dsp
\rarrow V_j + 2\alpha_j \phi_j V_j\\
\end{array}
\label{transform}
\end{eqnarray}
under which the spins $\vec S_j$ and the
Hamiltonian $H=\sum_j H_{int}[j]$ are invariant.
This is a ``supersymmetric'' transformation, generated by the $\phi_j$ fields
in the unitary 
transformation  $U=\prod_j(1 + \alpha_j \hat \phi_j)$.
Splitting the Berry phase term into  separate contributions from 
the spin (${\cal L}_B[\vec S]$) and $\phi$ field, 
\begin{equation}
\begin{array}{rl}
{1 \over 2} \vec \eta \partial_{\tau} \vec \eta
\equiv &\dsp {\cal L}_B[\vec S] + {1 \over 2} \phi \partial_{\tau} \phi\\
\rarrow &\dsp {\cal L}_B[\vec S] + {1 \over 2} (\phi+ \alpha) \partial_{\tau}
 (\phi+ \alpha) \\
\end{array}
\end{equation}
we find that this term transforms as 
\begin{equation}
\int d \tau \bigl[
{1 \over 2} \vec \eta_j \partial_{\tau} \vec \eta_j+\dot \alpha_j \phi_j 
\bigr]
\rarrow
\int d \tau\bigl[
{1 \over 2} \vec \eta_j \partial_{\tau} \vec \eta_j
- {1 \over 2} \dot \alpha_j{1 \over \partial_{\tau}}\dot\alpha_j \bigr]
\end{equation}
By differentiating
with respect to the source fields, we confirm that
the $\phi$ field is decoupled, with
a free propagator $\la \phi_j(1) \phi_j(2)\ra
= {1 \over \partial \tau} \equiv {1 \over 2}{\rm sgn} (\tau_1 - \tau_2)$.

These connections between  three-body bound-state formation and
odd-frequency pairing provide insight
into its possible application to heavy fermion physics.\cite{previous}
Coherence factors that 
are linear in  excitation
energy $\la \vk \vert \rho \vert \vk\ra\propto
\la \vk \vert S_z\vert \vk\ra\propto\omega$
and the development
of a quadratic temperature or frequency 
dependence in a wide variety of response functions, such as
the normalized NMR relaxation rate ${1 \over T_1T}$, the
transverse ultrasound attenuation $\alpha_T$,
the depletion of the superfluid density $\Delta\rho_s$ 
and  the quasiparticle conductivity $\si (\om)$, 
\begin{equation}
\left(
{1 \over T_1 T},  {\alpha_T},
\Delta\rho_s(T), \sigma(\omega) \right) \propto {\rm max}( T^2, \om^2)
\end{equation}
are seen as a  necessary consequence of 
neutral bound-states. 
These features are expected to co-exist with 
a linear specific heat and an essentially isotropic
thermal conductivity.
Crystalline
anisotropy, not included in our discussion, 
may influence bound state formation, for the
local bound-states should
acquire
a {\sl specific} crystal field symmetry $\Gamma$ of the lattice.
The  three-body spinor will take  the form
$
\xi_{\Gamma j}=
(\vec \si \cdot \vec S_j) \psi_{\Gamma j}
$
where $\psi_{\Gamma j} = \sum_{\vk} \gamma_{\Gamma \vk}
e^{i \vec k \cdot \vec R_j}\psi_{\vk}$ is a conduction
spinor with symmetry $\Gamma$. Hybridization 
with conduction electrons will reflect the same symmetry 
$V_{\vk} = V \gamma_{\Gamma \vk}$.
Nodes of the microscopic crystal field
f-wave function of the local moment can thus
lead to gap zeroes 
$\Delta_{\vk} \propto \gamma_{\Gamma \vk}^2$,
suggesting an interesting 
possibility of a co-existence between neutral
Fermi surfaces and gap lines of conventional quasiparticles.

We have attempted to elucidate the 
physics of odd-frequency superconductivity with the proposal
that it is driven by the
formation of neutral three-body bound states.  This hypothesis provides
a physical interpretation of the appearance of Majorana spin excitations
in the Kondo lattice model, and may be useful in developing
our understanding of heavy fermion superconductors.

We would like to thank E. Abrahams, P. W. Anderson, N. Andrei, 
L. Ioffe, G. Lonzarich, D. MacLaughlin, A. Ramirez and A. E. Ruckenstein
for discussions
related to this work.  This work was  supported by  NSF grant
DMR-93-12138, a grant from the SERC, UK and CNPq, Brazil.

\gdef\journal#1, #2, #3, 1#4#5#6{       
    {\sl #1~}{\bf #2}, #3 (1#4#5#6)}

\def\pr{\journal Phys. Rev., }

\def\pra{\journal Phys. Rev. A, }

\def\prb{\journal Phys. Rev. B, }

\def\prc{\journal Phys. Rev. C, }

\def\prd{\journal Phys. Rev. D, }

\def\prl{\journal Phys. Rev. Lett., }

\def\jmp{\journal J. Math. Phys., }

\def\rmp{\journal Rev. Mod. Phys., }

\def\cmp{\journal Comm. Math. Phys., }

\def\adv{\journal Adv. Phys., }

\def\ap{\journal Adv. Phys., }

\def\np{\journal Nucl. Phys., }

\def\pl{\journal Phys. Lett., }

\def\apj{\journal Astrophys. Jour., }

\def\apjl{\journal Astrophys. Jour. Lett., }

\def\jpc{\journal J. Phys. C, }

\def\jetp{\journal Sov. Phys. JETP, }

\def\jetl{\journal Sov. Phys. JETP Letters, }

\def\phil{\journal Philos. Mag.}

\def\ssc{\journal Solid State Commun., }

\def\annp{\journal Ann. Phys. (N.Y.), }

\def\zpb{\journal Zeit. Phys. B., }

\def\jdp{\journal J.  Phys. (Paris), }

\def\jdc{\journal J.  Phys. (Paris) Colloq., }

\def\jpjap{\journal J. Phys. Soc. Japan, }

\def\physica{\journal Physica B, }

\def\phl{\journal Phil. Mag.,}

\def\phb{\journal Physica B, }

\def\jmmm{\journal J. Magn. Mag. Mat., }

\def\jmm{\journal  J. Mag. Mat., }

\def\sci{\journal Science, }

\end{document}

%% file: defs.tex
\def\eps{\epsilon}
\def\3he{{$^3${\rm He}}}

\def\slD{\raise.15ex\hbox{$/$}\kern-.57em\hbox{$D$}}
\def\dsl{\raise.15ex\hbox{$/$}\kern-.57em\hbox{$\Delta$}}
\def\slp{{\raise.15ex\hbox{$/$}\kern-.57em\hbox{$\partial$}}}
\def\nsl{\raise.15ex\hbox{$/$}\kern-.57em\hbox{$\nabla$}}
\def\sla{\raise.15ex\hbox{$/$}\kern-.57em\hbox{$\rightarrow$}}
\def\slla{\raise.15ex\hbox{$/$}\kern-.57em\hbox{$\lambda$}}
\def\gtwid{\raise.3ex\hbox{$>$\kern-.75em\lower1ex\hbox{$\sim$}}}
\def\ltwid{\raise.3ex\hbox{$<$\kern-.75em\lower1ex\hbox{$\sim$}}}

\def\12{{1\over2}}

\def\part{\partial}
\def\la{\lambda}

\def\be{{\beta}}

\def\bethlogo{\vbox{\bf \line{\hrulefill} 
    \kern-.5\baselineskip 
    \line{\hrulefill\phantom{ ELIZABETH A. MASON }\hrulefill} 
    \kern-.5\baselineskip 
    \line{\hrulefill\hbox{ ELIZABETH A. MASON }\hrulefill} 
    \kern-.5\baselineskip 
    \line{\hrulefill\phantom{ 1411 Chino Street }\hrulefill} 
    \kern-.5\baselineskip 
    \line{\hrulefill\hbox{ 1411 Chino Street }\hrulefill} 
    \kern-.5\baselineskip 
    \line{\hrulefill\phantom{ Santa Barbara, CA 93101 }\hrulefill} 
    \kern-.5\baselineskip 
    \line{\hrulefill\hbox{ Santa Barbara, CA 93101 }\hrulefill}
    \kern-.5\baselineskip 
    \line{\hrulefill\phantom{ (805) 962-2739 }\hrulefill} 
    \kern-.5\baselineskip 
    \line{\hrulefill\hbox{ (805) 962-2739 }\hrulefill}}}
\def\lisalogo{\vbox{\bf \line{\hrulefill} 
    \kern-.5\baselineskip 
    \line{\hrulefill\phantom{ LISA R. GOODFRIEND }\hrulefill} 
    \kern-.5\baselineskip 
    \line{\hrulefill\hbox{ LISA R. GOODFRIEND }\hrulefill} 
    \kern-.5\baselineskip 
    \line{\hrulefill\phantom{ 6646 Pasado }\hrulefill} 
    \kern-.5\baselineskip 
    \line{\hrulefill\hbox{ 6646 Pasado }\hrulefill} 
    \kern-.5\baselineskip 
    \line{\hrulefill\phantom{ Santa Barbara, CA 93108 }\hrulefill} 
    \kern-.5\baselineskip 
    \line{\hrulefill\hbox{ Santa Barbara, CA 93108 }\hrulefill}
    \kern-.5\baselineskip 
    \line{\hrulefill\phantom{ (805) 962-2739 }\hrulefill} 
    \kern-.5\baselineskip 
    \line{\hrulefill\hbox{ (805) 962-2739 }\hrulefill}}}
\def\ka{{\kappa}}

\def\la{{\lambda}}

\def\low#1{\lower.5ex\hbox{${}_#1$}}
\def\ltwid{\raise.3ex\hbox{$<$\kern-.75em\lower1ex\hbox{$\sim$}}}

\def\om{{\omega}}

\def\psl{\raise.15ex\hbox{$/$}\kern-.57em\hbox{$\partial$}}
\def\partt{\raise.15ex\hbox{$\widetilde$}{\kern-.37em\hbox{$\partial$}}}
\def\parts{\raise.15ex\hbox{$/$}{\kern-.6em\hbox{$\partial$}}}
\def\nablas{\raise.15ex\hbox{$/$}{\kern-.6em\hbox{$\nabla$}}}
\def\oprod{\hbox{$\rm O$}{\kern -0.8em\hbox{$\Pi$}}}
\def\partw#1{\raise.15ex\hbox{$/$}{\kern-.6em\hbox{${#1}$}}}

\def\si{{\sigma}}

\def\gtappr{{{\lower4pt\hbox{$>$} } \atop \widetilde{ \ \ \ }}}
\def\ltappr{{{\lower4pt\hbox{$<$} } \atop \widetilde{ \ \ \ }}}

\def\topppageno1{\global\footline={\hfil}\global\headline
={\ifnum\pageno<\firstpageno{\hfil}\else{\hss\twelverm --\ \folio
\ --\hss}\fi}}

\def\toppageno2{\global\footline={\hfil}\global\headline
={\ifnum\pageno<\firstpageno{\hfil}\else{\rightline{\hfill\hfill
\twelverm \ \folio
\ \hss}}\fi}}

\def\joinrelde{\mathrel{\mkern-11mu}}
\def\joinrel{\mathrel{\mkern-9mu}}
\def\joinrelw{\mathrel{\mkern-6mu}}

\def\relbd{\mathrel{{\bf\smash{{\phantom- \above1pt \phantom-
}}}}}

\def\ltdash{\raise-1.8pt\hbox{$\scriptscriptstyle |$}}

%% file: defs2.tex
\def \ra{\rangle}
\def\la{\langle}

\def\dg{{^
{\dag}}}

\def\ra{\rangle}
\def\la{\langle}

\def\1{{\bf 1}}
\def\2{{\bf 2}}

\def\rarrow{\rightarrow}

\def\ul{\underline}

\def\vk{\vec k}

\def\ell{{\it l } {\rm n}}

\def\si{\sigma}

\def\cx2{\sqrt{c^2_x+c^2_y}}

\def\gkk{\gamma _{\vec k}}
\def\gk2{\gkk ^2}
\def\dw{\downarrow}
\def\up{\uparrow}
\def\gtappr{{{\lower4pt\hbox{$>$} } \atop \widetilde{ \ \ \ }}}
\def\ltappr{{{\lower4pt\hbox{$<$} } \atop \widetilde{ \ \ \ }}}

\def\dsp{\displaystyle}
\def\pbar{{\partial\kern-1.2ex\raise0.25ex\hbox{/}}}

\def\up{\uparrow}
\def\dw{\downarrow}

\def\dsp{\displaystyle}

\def\dg{{^{\dag}}}

\def\ra{\rangle}
\def\la{\langle}

\def\1{{\bf 1}}
\def\2{{\bf 2}}

\def\rarrow{\rightarrow}

\def\ul{\underline}

\def\vk{\vec k}

\def\ell{{\it l } {\rm n}}

\def\si{\sigma}

\def\cx2{\sqrt{c^2_x+c^2_y}}

\def\gkk{\gamma _{\vec k}}
\def\gk2{\gkk ^2}
\def\gtappr{{{\lower4pt\hbox{$>$} } \atop \widetilde{ \ \ \ }}}
\def\ltappr{{{\lower4pt\hbox{$<$} } \atop \widetilde{ \ \ \ }}}

\def\inu{i \nu_n}

\def\thickra{\hbox{\raise0.2pt\hbox{{$\bf >\mkern-13mu>\mkern-13mu>$}}}}
\def\thickrarrow{\hbox{\raise0.28pt\hbox{{$\bf >\mkern-13mu>\mkern-13mu>$}}}}

%% file: three.bbl
\begin{references}
\bibitem{note}Elementary defects of charge $e$ spinors 
correspond to a phase change
of $\pi$, and are associated with a flux quantum $(\hbar/e) \pi= h/2e$.
\bibitem{details} For odd $N$, using the property
$\prod_a S^a = i^{(N-1)/2}/2^N$,
we may invert the transformation,
$\phi_j = (-2i)^{(N-1)\over 2} \prod_a \eta_a[j]$, 
$\vec S_j = \phi_j \vec \eta_j$. For 
$N=3$ this recovers
the Majorana spin representation 
$\vec S_j = -(i/2) \vec \eta_j \times \vec \eta_j$.
\end{references}
